\documentclass[pdflatex,sn-mathphys-num]{sn-jnl}
\usepackage{graphicx}%
\usepackage{multirow}%
\usepackage{amsmath,amssymb,amsfonts}%
\usepackage{amsthm}%
\usepackage{mathrsfs}%
\usepackage[title]{appendix}%
\usepackage{xcolor}%
\usepackage{textcomp}%
\usepackage{manyfoot}%
\usepackage{booktabs}%
\usepackage{algorithm}%
\usepackage{algorithmicx}%
\usepackage{algpseudocode}%
\usepackage{listings}%

\begin{document}

\title{Quantum action of the Josephson dynamics}

\author*[1,2]{\fnm{Cesare} \sur{Vianello}}
\email{cesare.vianello@phd.unipd.it}

\author[1]{\fnm{Sofia} \sur{Salvatore}}

\author[1,2,3,4]{\fnm{Luca} \sur{Salasnich}}

\affil[1]{\orgdiv{Dipartimento di Fisica e Astronomia ``Galileo Galilei''}, \orgname{Università di Padova}, \orgaddress{\street{Via Marzolo 8}, \postcode{I-35131}, \city{Padova}, \country{Italy}}}

\affil[2]{\orgname{INFN Sezione di Padova}, \orgaddress{\street{Via Marzolo 8}, \postcode{I-35131}, \city{Padova}, \country{Italy}}}

\affil[3]{\orgdiv{Padua QTech Center}, \orgname{Università di Padova}, \orgaddress{\street{Via Gradenigo 6/A}, \postcode{I-35131}, \city{Padova}, \country{Italy}}}

\affil[4]{\orgname{CNR-INO}, \orgaddress{\street{Via Carrara 1}, \postcode{I-50019}, \city{Sesto Fiorentino}, \country{Italy}}}

\abstract{We study the beyond-mean-field Josephson dynamics of the relative phase between two coupled macroscopic quantum systems. Using a covariant background field method, we derive the one-loop only-phase quantum effective action and the corresponding equation of motion for the quantum average of the phase. These analytical results are benchmarked against the exact quantum dynamics of the two-site Bose-Hubbard model, demonstrating a relevant improvement over the standard mean-field predictions across a wide range of interaction strengths.}


\maketitle

\section{Introduction}

Superfluids and superconductors are macroscopic quantum states of matter that can be described by a single wavefunction. If two such systems are coupled together, particle currents should oscillate between them, accompanied by oscillations of the relative phase between the two macroscopic wavefunctions. This phenomenon was originally predicted by Josephson in 1962 for superconductors \cite{Josephson} and observed soon after \cite{Anderson, Shapiro}. Since then, the Josephson effect has been studied both theoretically and experimentally in a variety of coupled macroscopic quantum systems, such as superfluid helium reservoirs \cite{Pereverzev, Sukhatme}, double well trapped Bose-Einstein condensates \cite{Smerzi, Albiez, Spagnolli}, dipolar condensates \cite{Xiong, Abad, Adhikari, Allende}, momentum-space condensates \cite{Mukhopadhyay}, magnon condensates \cite{Nakata, Kreil}, polariton condensates \cite{Lagoudakis, Abbarchi}, supersolids \cite{Biagioni, Alana, donelli}, and fermionic superfluids \cite{Burchianti, Luick, Valtolina, Pascucci}.

Due to the coherence of the systems considered, the Josephson dynamics is often described in a mean-field approximation. In the context of two weakly linked condensates, denoted $L$ and $R$, the mean-field equations for the time evolution of the relative phase $\phi \equiv \phi_R-\phi_L$ and the population imbalance $z \equiv (N_L-N_R)/N$ are
\begin{subequations}\label{eqss}
\begin{align}
    \hbar\dot\phi &= \frac{2Jz}{\sqrt{1-z^2}}\cos\phi + UNz + \varepsilon,\label{eqss1}\\
    \hbar\dot z &= -2J\sqrt{1-z^2}\sin\phi,\label{eqss2}
\end{align}
\end{subequations}
where $U$ is the boson-boson interaction, $J>0$ is the tunneling energy, $\varepsilon$ is a potential energy difference between the two condensates, and $N$ is the total number of particles \cite{Smerzi}. A simplified version of these equations in the regime $|z|\ll 1$, namely $\dot\phi=\varepsilon/\hbar$ and $\dot z=-(2J/\hbar)\sin\phi$, govern the dynamics of superconductive Josephson junctions, where $\dot z$ is proportional to the electric current and $\varepsilon$ to the applied potential difference across the junction \cite{Josephson}.

In this paper, we address the role of quantum fluctuations and evaluate first-order quantum corrections to the mean-field dynamics described by Eqs. \eqref{eqss} focusing on the relative phase $\phi$, which is directly related to the coherence of the system \cite{Vianello}. Throughout the work we will consider $U>0$, which corresponds to repulsive interaction, and $\varepsilon=0$. We start from the observation that Eqs. \eqref{eqss} (with $\varepsilon=0$) are the Euler-Lagrange equations for the action
\begin{equation}\label{action}
    S[\phi, z] = \int dt\left[\frac{N\hbar z}{2} \dot \phi - \frac{UN^2}{4}z^2 + JN\sqrt{1-z^2}\cos\phi\right]
\end{equation}
with the conserved energy
\begin{equation}\label{ener}
    E(\phi, z) = \frac{UN^2}{4}z^2 - JN\sqrt{1-z^2}\cos\phi.
\end{equation}
In particular, linearizing around the equilibrium $\phi=z=0$, one finds that small oscillations of $\phi$ and $z$ occur with the frequency
\begin{equation}\label{Jf}
    \omega_J = \frac{\sqrt{2J(UN+2J)}}{\hbar}.
\end{equation}
All quantum effects are contained in the transition amplitude between two states parametrized by the variables $(\phi, z)$, which in the path integral approach reads $\int \mathcal D\phi\,\mathcal Dz\,\exp(\frac{i}{\hbar}S[\phi,z])$, and of which the mean-field equations \eqref{eqss} are the stationary-phase approximation. This quantum theory is in correspondence with the two-site Bose-Hubbard model
\begin{equation}\label{BH}
    \hat H = \frac{U}{2}\left[\hat N_L(\hat N_L-1)+\hat N_R(\hat N_R-1)\right] - J\left(\hat a_L^\dag \hat a_R + \hat a_R^\dag \hat a_L\right),
\end{equation}
where $\hat a^{(\dag)}_{L(R)}$ are bosonic creation and annihilation operators and $\hat N_{L(R)}\equiv\hat a^\dag_{L(R)}\hat a_{L(R)}$ are the corresponding number operators \cite{Leggett}. The total number $\hat N = \hat N_L+\hat N_R$ is conserved; for fixed $N$, the Hamiltonian is thus defined on the $(N+1)$-dimensional Hilbert space spanned by the Fock basis $\{|N_L, N_R\rangle\} = \{|j, N-j\rangle\}_{j=0,\dots,N}$. The transition amplitudes of $\hat H$ can be written in the basis of bosonic coherent states $|\alpha\rangle\equiv |a_L\rangle\otimes|a_R\rangle$, that are eigenstates of $\hat a_{L(R)}$ with eigenvalues $a_{L(R)} = \sqrt{N_{L(R)}}e^{i\phi_{L(R)}}$, as
\begin{align}
    \langle \alpha_f|e^{-i\hat H (t_f-t_i)/\hbar}|\alpha_i\rangle &= \int\limits_{a_{L}(t_i)=a_{Li}}^{a^*_{L}(t_f)=a^*_{Lf}} \mathcal D[a_L^*, a_L] \int\limits_{a_{R}(t_i)=a_{Ri}}^{a^*_{R}(t_f)=a^*_{Rf}} \mathcal D[a_R^*,a_R]\nonumber\\
    & \times e^{N_f} e^{\frac{i}{\hbar}\int_{t_i}^{t_f} dt\,\Bigl[i\hbar(a_L^*\dot a_L + a_R^*\dot a_R)-H(a_L^*,a_L, a_R^*,a_R)\Bigr]}\nonumber\\
    &= N_f e^{N_f}\delta(N_f-N_i)\int\limits_{\phi(t_i)=\phi_i}^{\phi(t_f)=\phi_f}\mathcal D\phi\int\limits_{z(t_i)=z_i}^{z(t_f)=z_f}\mathcal Dz\,e^{\frac{i}{\hbar}S[\phi,z]},
\end{align}
where the path integral is restricted to trajectories in $a$-space for which the average number of particles $N=N_L+N_R$ is time-independent and fixed by the boundary conditions \cite{Negele, Furutani}. This establishes the correspondence between the operational approach based on $\hat H$ and the path integral approach based on $S[\phi,z]$. The mean-field equations \eqref{eqss} can be derived equivalently either from the stationary-phase approximation of the path integral or by averaging the Heisenberg equations generated by $\hat H$ over bosonic coherent states.

In the following, we compute analytically the first-order quantum corrections to the dynamics of the phase by first deriving an effective action for $\phi$ integrating out $z$ at the Gaussian level (Section \ref{Sec2}) and then computing the corresponding one-loop quantum effective action (Section \ref{Sec3}), generalizing a treatment previously presented in Ref. \cite{Furutani}. The range of validity of our results, in terms of interaction and number of particles, will be discussed by comparing them with fully quantum numerical results obtained by exact diagonalization of the Bose-Hubbard Hamiltonian (Section \ref{Sec4}).

\section{Only-phase effective action}\label{Sec2}

Given the action $S[\phi,z]$, the effective action for the phase $\mathcal A[\phi]$ is defined as \cite{Burgess}
\begin{equation}\label{wilson}
    e^{\frac{i}{\hbar}\mathcal A[\phi]} = \int \mathcal Dz\,e^{\frac{i}{\hbar}S[\phi,z]}.
\end{equation}
The path integral can be computed explicitly expanding  $S[\phi, z]$ up to second order around $z = 0$. This perturbative step assumes simultaneously that $E(\phi,z)$ is well approximated by its quadratic part in $z$ near $z=0$ and that during the dynamics $z(t)$ remains confined to that small-$z$ region, so as to never probe the anharmonic terms in $E(\phi,z)$. Since $E(\phi,z) \simeq \frac{N}{4}[(UN+2J)z^2 + \frac{J}{2}z^4\cos\phi]-JN\cos\phi$ and $z \simeq \hbar\dot\phi/(UN+2J\cos\phi)$ around $z=0$, both conditions are satisfied if $\Lambda \equiv UN/2J \gg 1$. An actual estimate of the lower bound of $\Lambda$ will be provided in Section \ref{Sec4}. In this way $S[\phi,z] \simeq S^{(2)}[\phi, z]$, where
\begin{align}\label{Sexp}
    S^{(2)}[\phi,z] = \int dt\biggl[- \frac{1}{2m(\phi)}\left(\frac{N\hbar z}{2}\right)^2+ \frac{N\hbar z}{2}\dot\phi-V(\phi)\biggr]
\end{align}
and
\begin{equation}\label{mV}
    m(\phi) = \frac{N\hbar^2}{2(UN+2J\cos\phi)},\qquad V(\phi)= -JN\cos\phi.
\end{equation}
Here $p_\phi \equiv N\hbar z/2$ plays the role of conjugate momentum of $\phi$, and $\int \mathcal D\phi \mathcal Dz \exp(\frac{i}{\hbar} S^{(2)}[\phi,z])$ has exactly the form one encounters when computing the phase-space path integral for a quantum particle with coordinate $\phi$ moving in a potential $V(\phi)$ \cite{kleinert}. In this case integrating out $z$ corresponds to the standard passage from the phase-space to the configuration-space path integral. Here, however, the spatial dependence of the mass makes this step nontrivial, for the action $S^{(2)}[\phi,z]$ describes a quantum particle in a curved space with metric $g_{\mu\nu} = m(\phi)\eta_{\mu\nu}$ \cite{Bastianelli}. Completing the square by shifting the momentum as $\tilde p_\phi = p_\phi-m(\phi)\dot\phi$ ($\mathcal D\tilde p_\phi=\mathcal Dp_\phi$), we get
\begin{equation}\label{cea}
    e^{\frac{i}{\hbar}\mathcal A[\phi]} = e^{\frac{i}{\hbar} \int dt\,[\frac{m(\phi)}{2}\dot\phi^2-V(\phi)]}\int \mathcal D\tilde p_\phi\,e^{-\frac{i}{\hbar} \int dt\,\frac{\tilde p_\phi^2}{2m(\phi)}}.
\end{equation}
By time slicing, the remaining Gaussian integral is $\prod_j \int \frac{d\tilde p_{\phi j}}{2\pi\hbar} \exp(-\frac{i}{\hbar}\delta t \frac{\tilde p_{\phi j}^2}{2m(\phi_j)}) = \prod_j\sqrt{m(\phi_j)/2\pi\hbar i\delta t}$, and would contribute to $\mathcal A[\phi]$ with a term proportional to $\sum_j \ln m(\phi_j)$, which in the continuum limit $\frac{1}{\delta t}\sum_j \ln m(\phi_j)\delta t \to \delta(0)\int dt \ln m(\phi(t))$ is divergent. In the mode regularization scheme \cite{Bastianelli, kleinert}, the divergence is reabsorbed by defining the invariant path integral measure as
\begin{equation}\label{measure}
    \mathcal D\mu(\phi) \equiv \frac{1}{Z}\prod_j d\phi_j\sqrt{m(\phi_j)},
\end{equation}
where $Z = \prod_j \sqrt{2\pi\hbar i \delta t}$. 

Thanks to the nontrivial measure, the effective action takes the form
\begin{equation}\label{only_phase}
    \mathcal A[\phi] = \int dt\left[\frac{m(\phi)}{2}\dot\phi^2 -V(\phi)\right].
\end{equation}
This is simply the tree-level action one obtains by substituting for $z$ in Eq. \eqref{Sexp} the solution $z = z_\text{cl}(t)$ of the equations of motion. Indeed, writing $z = z_\text{cl}+\tilde z$ and expanding $S^{(2)}[\phi,z]$ in the fluctuations $\tilde z$, we get $S^{(2)}[\phi,z] = S^{(2)}[\phi, z_\text{cl}] + \frac{1}{2}\int dt\int dt'\,\tilde z(t)\frac{\delta^2S^{(2)}[\phi, z_\text{cl}]}{\delta z(t)\delta z(t')}\tilde z(t') = S^{(2)}[\phi, z_\text{cl}] - \int dt\frac{(N\hbar\tilde z/2)^2}{2m(\phi)}$, so that $\exp(\frac{i}{\hbar}\mathcal A[\phi]) = \exp(\frac{i}{\hbar}S^{(2)}[\phi, z_\text{cl}])\int \mathcal D\tilde z\,\exp(-\frac{i}{\hbar}\int dt\,\frac{(N\hbar\tilde z/2)^2}{2m(\phi)})$. Comparing this with Eq. \eqref{cea}, we see that $\mathcal A[\phi] = S^{(2)}[\phi, z_\text{cl}]$. 

\section{Quantum effective action}\label{Sec3}

Quantum corrections to the dynamics of $\mathcal A[\phi]$ can be computed within the quantum effective action formalism \cite{Burgess, kleinert}, which was introduced many years ago to study quantum effects on field theories with spontaneously broken symmetries \cite{Goldstone, Jona, Coleman, Jackiw}. We recall that the quantum effective action is defined as the Legendre transform ${\Gamma[\Phi] = W[J]-\int dt\,J\Phi}$ of the generating functional of connected correlation functions, $W[J] = -i\hbar \ln Z[J] = -i\hbar \ln\int \mathcal D\mu(\phi) \exp [\frac{i}{\hbar}\int dt\,(\mathcal A[\phi] + J \phi)]$, where $\Phi = \delta W[J]/\delta J$ is the quantum average of $\phi$ in the presence of the source $J$. Since $\delta\Gamma[\Phi]/\delta\Phi = -J$, the quantum average $\Phi$ in absence of external sources extremizes $\Gamma[\Phi]$; this is the principle of least action for the full quantum theory. Consequently, the sum of connected diagrams built from the classical action $\mathcal A[\phi]+\int dt\,J\phi$ can be obtained from the tree diagrams of the effective action $\Gamma[\Phi] + \int dt\,J\Phi$. In terms of the fluctuations $\eta \equiv \phi - \Phi$, we thus have
\begin{equation}
    e^{\frac{i}{\hbar}\Gamma[\Phi]} = \int \mathcal D\mu(\eta)\,e^{\frac{i}{\hbar}(\mathcal A[\Phi+\eta]-\int dt\,\eta\,\delta\Gamma[\Phi]/\delta\Phi)},
\end{equation}
where the measure $\mathcal D\mu(\eta)$ is defined according to Eq. \eqref{measure} as $\mathcal D\mu(\eta) = \frac{1}{Z}\prod_j d\eta_j \sqrt{m(\Phi+\eta_j)} = \frac{1}{Z}\prod_j d\eta_j \sqrt{m(\Phi)}\,e^{\frac{1}{2}\delta(0)\int dt\,\ln\frac{m(\Phi+\eta)}{m(\Phi)}} \equiv \mathcal D\eta\sqrt{m(\Phi)}$.

The background field method consists in computing $\Gamma[\Phi]$ perturbatively by expanding $\mathcal A[\Phi+\eta]$ in powers of $\eta$. At the one-loop level we have $\Gamma[\Phi] = \mathcal A[\Phi] + \Gamma_1[\Phi] + \mathcal O(\hbar^2)$, where
\begin{align}
    \Gamma_1[\Phi] = -i\hbar \ln \int \mathcal D\eta\sqrt{m(\Phi)}\,e^{\frac{i}{\hbar}\mathcal A^{(2)}[\Phi,\eta]}\label{1lqea}
\end{align}
and $\mathcal A^{(2)}[\Phi,\eta]$ is the quadratic term of the expansion of $\mathcal A[\Phi+\eta]$. Due to the $\Phi$ dependence of the mass, this expansion must be done covariantly to ensure that the quantum effective action remains manifestly invariant under changes of coordinate. Thus following Ref. \cite{Kleinert_PLA},
\begin{equation}
    \mathcal A^{(2)}[\Phi,\eta] = \frac{1}{2}\int dt\,dt'\,\eta(t)\frac{D^2\mathcal A[\Phi]}{\delta\phi(t)\delta\phi(t')}\eta(t'),
\end{equation}
where $D/\delta\phi$ denotes the covariant functional derivative. In particular, the second covariant derivative is given by
\begin{equation}
    \frac{D^2\mathcal A[\Phi]}{\delta\phi(t)\delta\phi(t')} = \frac{\delta^2\mathcal A[\Phi]}{\delta\phi(t)\delta\phi(t')} - \gamma(\Phi(t))\frac{\delta \mathcal A[\Phi]}{\delta\phi(t')},
\end{equation}
where $\gamma(\Phi) = m'(\Phi)/2m(\Phi)$ is the one-dimensional Christoffel symbol for the metric $g_{\mu\nu}(\Phi) = m(\Phi)\eta_{\mu\nu}$, and
\begin{gather}
    \frac{\delta\mathcal A[\Phi]}{\delta\phi(t)} = -V'(\Phi) - \frac{1}{2}m'(\Phi)\dot\Phi^2 - m(\Phi)\ddot\Phi,\\
    \frac{\delta^2\mathcal A[\Phi]}{\delta\phi(t)\delta\phi(t')} = -\left[m(\Phi)\partial_t^2 + V''(\Phi) + m'(\Phi)\left(\ddot\Phi + \dot\Phi \partial_t\right) + \frac{1}{2}m''(\Phi)\dot\Phi^2\right]\delta(t-t').
\end{gather}
Introducing the new coordinate $\tilde\eta = \sqrt{m(\Phi)}\eta$, we obtain
\begin{align}
    \eta(t)\frac{D^2\mathcal A[\Phi]}{\delta\phi(t)\delta\phi(t)}\eta(t) = \tilde\eta(t) \left[-\partial_t^2-\Omega^2(\Phi)\right]\tilde\eta(t),
\end{align}
where
\begin{equation}\label{omega2}
    \Omega^2(\Phi) = \frac{V''(\Phi)-\gamma(\Phi)V'(\Phi)}{m(\Phi)}.
\end{equation}
Since $\mathcal D\tilde\eta = \mathcal D\eta\sqrt{m(\Phi)}$, we conclude that the path integral in Eq. \eqref{1lqea} is equal to
\begin{equation}\label{gamma1}
    \Gamma_1[\Phi] = \frac{i\hbar}{2}\text{Tr} \ln \left[-\partial_t^2-\Omega^2(\Phi)\right].
\end{equation}

\begin{figure}
    \centering
    \includegraphics[width=\linewidth]{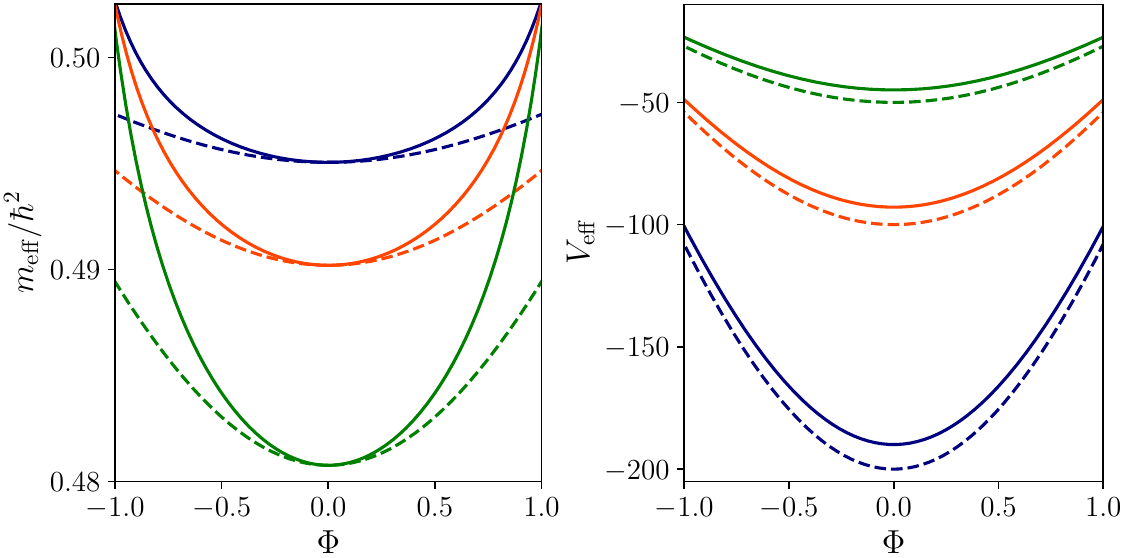}
    \caption{Effective mass (left panel) and effective potential as functions of $\Phi$ for $U=J=1.0$ and $N=50$ (green lines), 100 (orange lines), and 200 (blue lines) [Eqs. \eqref{meff}-\eqref{veff}]. The dashed lines represent the corresponding classical results [Eq. \eqref{mV}].}
    \label{fig1}
\end{figure}

The trace in Eq. \eqref{gamma1} involves a nonlocal functional of $\Phi(t)$ and cannot be computed exactly. If $\Omega^2(\Phi)$ varies slowly in time, however, one can build a derivative expansion around a constant $\Phi$, which yields asymptotically a local expression for $\Gamma_1[\Phi]$ \cite{Kleinert_PLA, Fraser, Iliopoulos, Cametti}. Such adiabatic approximation is valid as long as $|\dot\Omega(\Phi)|/\Omega^2(\Phi) \ll 1$. One then finds that $\Gamma_1[\Phi]$ can be written as the time integral of a local Lagrangian, which is a series of terms involving time derivatives of $\Phi(t)$ of increasing order,
\begin{equation}\label{derexp}
    \Gamma_1[\Phi] = \int dt \left[-V_{1}(\Phi) + \frac{Z_1(\Phi)}{2}\dot\Phi^2 + \mathcal O(\partial_t^4)\right].
\end{equation}
Here $V_1(\Phi)$ is the one-loop correction to the classical potential, and can be found by computing the trace in Eq. \eqref{gamma1} for a constant $\Phi$:
\begin{equation}
    V_{1}(\Phi) = \frac{\hbar\Omega(\Phi)}{2}.
\end{equation}
It represents the zero-point energy of Gaussian fluctuations with frequency $\Omega(\Phi)$. The one-loop correction to the mass, $Z_1[\Phi]$, can then be found by expanding $\Phi(t)$ in Eqs. \eqref{gamma1} and \eqref{derexp} around a constant value and matching the two expressions. This yields
\begin{equation}
    Z_1(\Phi) = \frac{\hbar}{32}\frac{[\partial_\Phi\Omega^2(\Phi)]^2}{\Omega^5(\Phi)}.
\end{equation}
The quantum effective action at one-loop and first order in the derivative expansion is therefore
\begin{equation}
    \Gamma[\Phi] = \mathcal A[\Phi] + \Gamma_1[\Phi] = \int dt\left[\frac{m_\text{eff}(\Phi)}{2}\dot\Phi^2 - V_\text{eff}(\Phi)\right],
\end{equation}
where
\begin{equation}\label{mVeff}
    m_\text{eff}(\Phi) = m(\Phi) + \frac{\hbar}{32}\frac{[\partial_\Phi\Omega^2(\Phi)]^2}{\Omega^5(\Phi)},\qquad V_\text{eff}(\Phi) = V(\Phi) + \frac{\hbar\Omega(\Phi)}{2}
\end{equation}
are the effective mass and the effective potential, respectively. It follows that $\Phi(t)$ satisfies the equation of motion
\begin{equation}\label{eomeff}
    \ddot\Phi + \frac{m'_\text{eff}(\Phi)}{2m_\text{eff}(\Phi)}\dot\Phi^2+\frac{V_\text{eff}'(\Phi)}{m_\text{eff}(\Phi)} = 0.
\end{equation}
Given the expressions in Eq. \eqref{mV} for the mass and the potential of the classical action, from Eq. \eqref{omega2} we get $\Omega^2(\Phi) = 2J[UN\cos\Phi + J(3\cos^2\Phi-1)]/\hbar^2$,  which substituted into Eq. \eqref{mVeff} yields
\begin{align}
    m_\text{eff}(\Phi) &= \frac{N\hbar^2}{2(UN+2J\cos\Phi)} + \frac{\hbar^2}{32\sqrt{2J}}\frac{(UN + 6J\cos\Phi)^2\sin^2\Phi}{\left[UN\cos\Phi + J(3\cos^2\Phi-1)\right]^{5/2}},\label{meff}\\
    V_\text{eff}(\Phi) &= -JN\cos\Phi + \sqrt{\frac{J}{2}\left[UN\cos\Phi + J(3\cos^2\Phi-1)\right]}.\label{veff}
\end{align}
With these functions, represented in Fig. \ref{fig1}, Eq. \eqref{eomeff} describes the quantum-corrected Josephson dynamics of the relative phase between two coupled condensates. We notice that if $\Phi(t) \sim \Phi_0 \cos(\omega_J t)$, then $|\dot\Omega(\Phi)|/\Omega^2(\Phi) \sim |\Phi_0|$ as order of magnitude, therefore we expect the adiabatic approximation to hold as long as the amplitude of phase oscillations remains much smaller than unity.

Quantum corrections do not change the position of the minimum of the potential, which is still located at $\Phi=0$, where also $m'_\text{eff}(0)=0$. In particular, small oscillations around $\Phi = 0$ are harmonic, with the frequency
\begin{align}\label{omegacorr}
    \Omega_J &= \sqrt{\frac{V''_\text{eff}(0)}{m_\text{eff}(0)}} = \omega_J \sqrt{1-\frac{1}{2N}\frac{UN+6J}{\sqrt{2J(UN+2J)}}},
\end{align}
where $\omega_J$ is the classical frequency \eqref{Jf}. In the limit $UN \gg J$, where $m$ can be approximated to be constant, this reduces to
\begin{equation}
    \Omega_J \simeq \frac{\sqrt{2JUN}}{\hbar} \sqrt{1-\sqrt\frac{U}{8JN}},
\end{equation}
that is the result of Ref. \cite{Furutani}.

\section{Quantum dynamics of the phase}\label{Sec4}

It is valuable to compare the quantum-corrected Josephson dynamics of Eqs. \eqref{eomeff}-\eqref{veff} with the fully quantum dynamics of the Bose-Hubbard Hamiltonian $\hat H$ [Eq. \eqref{BH}]. In general, the quantum evolution depends sensitively on the initial state. To establish a meaningful comparison with a semiclassical dynamics of the collective variables $(\phi, z)$, the initial state should possess well-defined average relative phase and population imbalance. As these are conjugate variables, the natural choice is a state that minimizes the product of their variances. Therefore, we initialize the system at $t=0$ in the $N$-particle atomic coherent state
\begin{equation}
    |\Psi_{\phi_0,z_0}\rangle = \frac{1}{\sqrt{N!}}\biggl(\sqrt{\frac{1+z_0}{2}}\hat a_L^\dag + \sqrt{\frac{1-z_0}{2}}e^{i\phi_0}\hat a_R^\dag\biggr)^N |0,0\rangle,
\end{equation}
that is a minimal-uncertainty product state with relative phase $\phi_0$ and population imbalance $z_0$ \cite{Arecchi, Leggett, Wimberger}. For $U=0$, $|\Psi_{0,\frac{1}{2}}\rangle$ is the exact ground state of $\hat H$ \cite{Vianello}. This choice allows a direct comparison between the quantum evolution and the semiclassical trajectories for $(\phi, z)$ with initial conditions $\phi(0)=\phi_0$ and $z(0)=z_0$. For the quantum-corrected dynamics, Eq. \eqref{eomeff}, the corresponding initial conditions are $\Phi(0)=\phi_0$ and $\dot\Phi(0) = N\hbar z_0/2m_\text{eff}(\phi_0)$.

The state $|\Psi_{\phi_0,z_0}\rangle$ is written in the Fock basis as ${|\Psi_{\phi_0,z_0}\rangle = \sum_{j=0}^N A_j|j, N-j\rangle}$, where $A_j =\sqrt{\binom{N}{j}\frac{1}{2^N}}(1+z_0)^{j/2}(1-z_0)^{(N-j)/2} e^{i(N-j)\phi_0}$. Since the Fock states are related to the eigenstates $|E_n\rangle$ of $\hat H$ by $|j, N-j\rangle = \sum_{n=0}^N c_j^{(n)}|E_n\rangle$, where the coefficients $c_j^{(n)}$ are real and normalized to unity, $|\Psi_{\phi_0,z_0}\rangle$ evolves unitarily in time as
\begin{equation}
     |\Psi(t)\rangle = e^{-i\hat Ht/\hbar}|\Psi_{\phi_0,z_0}\rangle = \sum_{k=0}^N A_k(t)|k, N-k\rangle,
\end{equation}
where $A_k(t) = \sum_{j=0}^N A_j \sum_{n=0}^N c_j^{(n)}c_k^{(n)}e^{-iE_nt/\hbar}$. The system at time $t$ is thus described by the density matrix $\hat\rho(t) = |\Psi(t)\rangle\langle\Psi(t)|$. The one-body density matrix $\hat\rho^{(1)} = N\,\text{Tr}_{2,\dots,N}\,\hat\rho(t)$ is in this case the $2\times 2$ Hermitian matrix with elements
\begin{equation}
    \rho^{(1)}_{ij}(t) = \langle \Psi(t) | \hat a^\dag_i \hat a_j|\Psi(t)\rangle,\qquad i,j \in \{L,R\},
\end{equation}
Since $\text{Tr}\,\hat\rho^{(1)}=N$, it has two eigenvalues $\varrho_0(t)$ and $\varrho_1(t)=N-\varrho_0(t) \le \varrho_0(t)$. If the largest eigenvalue satisfies $\varrho_0(t)/N \gg 1/2$, the corresponding eigenvector
\begin{equation}
    |\chi(t)\rangle = \chi_L(t)|L\rangle+\chi_R(t)|R\rangle
\end{equation}
represents the BEC order parameter, with the components $\chi_{L(R)}(t)$ giving the amplitudes on the left and right condensate orbitals. The quantity $\varrho_0(t)/N$ is thus a measure of the system's coherence, and the relative phase between the two condensates is
\begin{equation}\label{phi_ex}
    \phi_\text{cond}(t) = \arg[\chi_R(t)]-\arg[\chi_L(t)].
\end{equation}
Since at $t=0$ the system is initialized in the pure condensate state $|\Psi_{\phi_0,z_0}\rangle$, we have $\varrho_0(0)/N = 1$. Under quantum evolution, the system generally departs from an atomic coherent state, resulting in a decrease of $\varrho_0(t)/N$. In order for $\phi_\text{cond}(t)$ to retain its physical meaning, $\varrho_0(t)/N$ must remain sufficiently above $1/2$, the value corresponding to complete incoherence.

\begin{figure}
    \centering
    \includegraphics[width=0.78\linewidth]{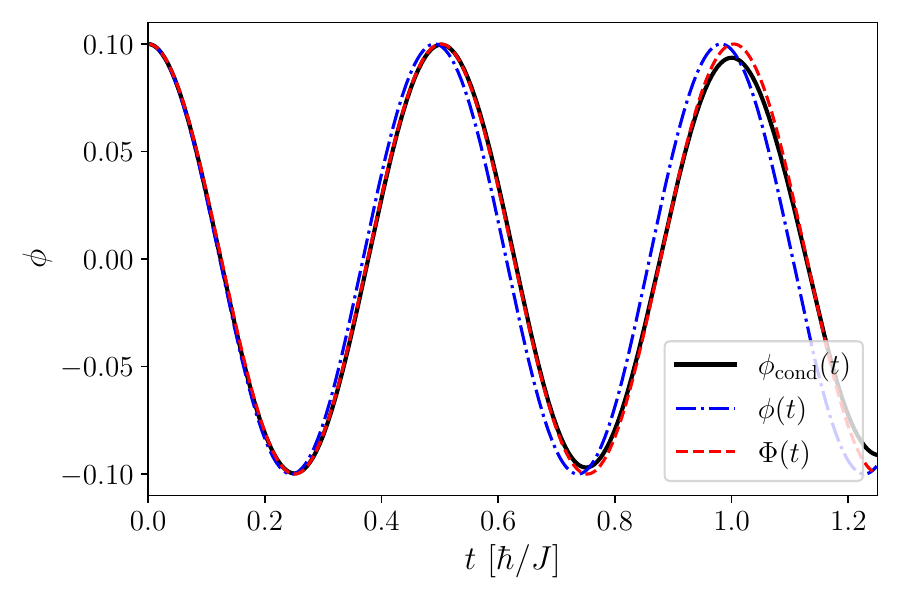}
    \caption{Comparison between the exact dynamics [Eq. \eqref{phi_ex}] (solid black line), the mean-field dynamics [Eq. \eqref{eqss}] (dashed-dotted blue line), and the quantum-corrected dynamics [Eqs. \eqref{eomeff}-\eqref{veff}] (dashed red line) of the relative phase, for $N=80$, $U=J=1.0$, $\phi_0=0.1$, and $z_0=0$. (Units: $\hbar=1$).}
    \label{fig2}
\end{figure}

To extract $\phi_\text{cond}(t)$, we first numerically diagonalize $\hat H$ to obtain the coefficients $\{c_k^{(n)}\}$. Since the dimension of the Hilbert space scales linearly with $N$, this can be easily done up to several thousand particles without the need for particularly sophisticated algorithms. We then construct the one-body density matrix and numerically diagonalize it at any given time, identifying the largest-eigenvalue eigenvector. In practice, we discretize the time interval of interest into $10^3$-$10^5$ steps. An example of the results we obtain is shown in Fig. \ref{fig2}, where the exact quantum dynamics is compared with the mean-field and the quantum-corrected dynamics (in the figure we use units where $\hbar=J=1$, so that times are also adimensional). For the chosen parameter values and initial conditions, $\langle\varrho_0(t)\rangle_\mathrm{time}/N \simeq 0.94$, and both semiclassical dynamics are well described by harmonic oscillations with frequencies $\omega_J$ and $\Omega_J$, respectively, with the quantum-corrected frequency being $2.1\%$ lower than the classical one. The harmonic approximation based on $\Omega_J$ remains accurate for oscillation amplitudes up to $0.3$, beyond which the full equation of motion \eqref{eomeff} is needed to accurately describe the dynamics. We see that the dynamics obtained from the quantum effective action significantly improves the comparison with the exact result, and the agreement is almost perfect until about two oscillation periods, when the amplitude modulation of the oscillations of $\phi_\text{cond}(t)$ starts to become evident. The latter is an expected, genuine many-body quantum effect \cite{Milburn, Sakmann}. In fact, $|\Psi(t)\rangle$ is a superposition of many eigenstates, so expectation values such as $\phi_\text{cond}(t)$ contain summations of terms like $e^{-i(E_n-E_m)t/\hbar}$. Because the energy levels of the interacting Hamiltonian are non-equally spaced, a band of closely spaced frequency differences appears and their superposition produces oscillations modulated by a slowly-varying envelope, showing beats, partial collapse and revivals. This mechanism is controlled by $U/J$ and $N$, with stronger interactions and smaller number of particles making the modulations faster and more pronounced (see also the discussion below). This behavior cannot be described by the quantum effective action, which is still a semiclassical description for the collective degrees of freedom of the system; nevertheless, it correctly captures the frequency of fast oscillations over several periods.

\begin{figure}
    \centering
    \includegraphics[width=0.78\linewidth]{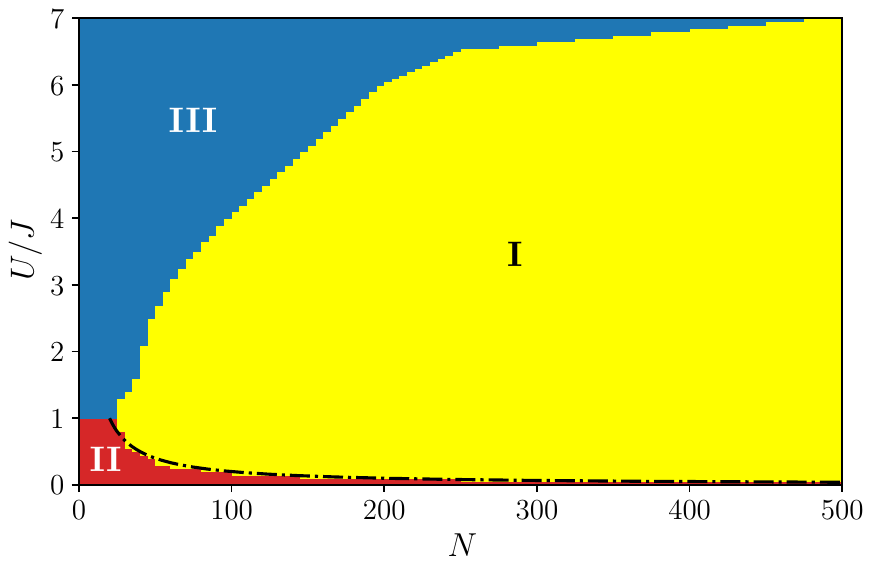}
    \caption{Range of validity of the quantum-corrected dynamics in terms of $U/J$ and $N$. In Region I (yellow) the quantum-corrected dynamics improves the mean field by bringing it closer to the exact dynamics. In Region II (red) the quantum-corrected dynamics performs worse than the mean field due to the breakdown of the initial Gaussian integration over $z$. In Region III (blue) the exact dynamics is dominated by strong anharmonicity, and neither the mean field nor the quantum-corrected dynamics provide an accurate description. The transition line between Regions I and II fits $\Lambda = UN/2J=10$ (black dashed-dotted line).}
    \label{fig3}
\end{figure}

\begin{figure}
    \centering
    \includegraphics[width=\linewidth]{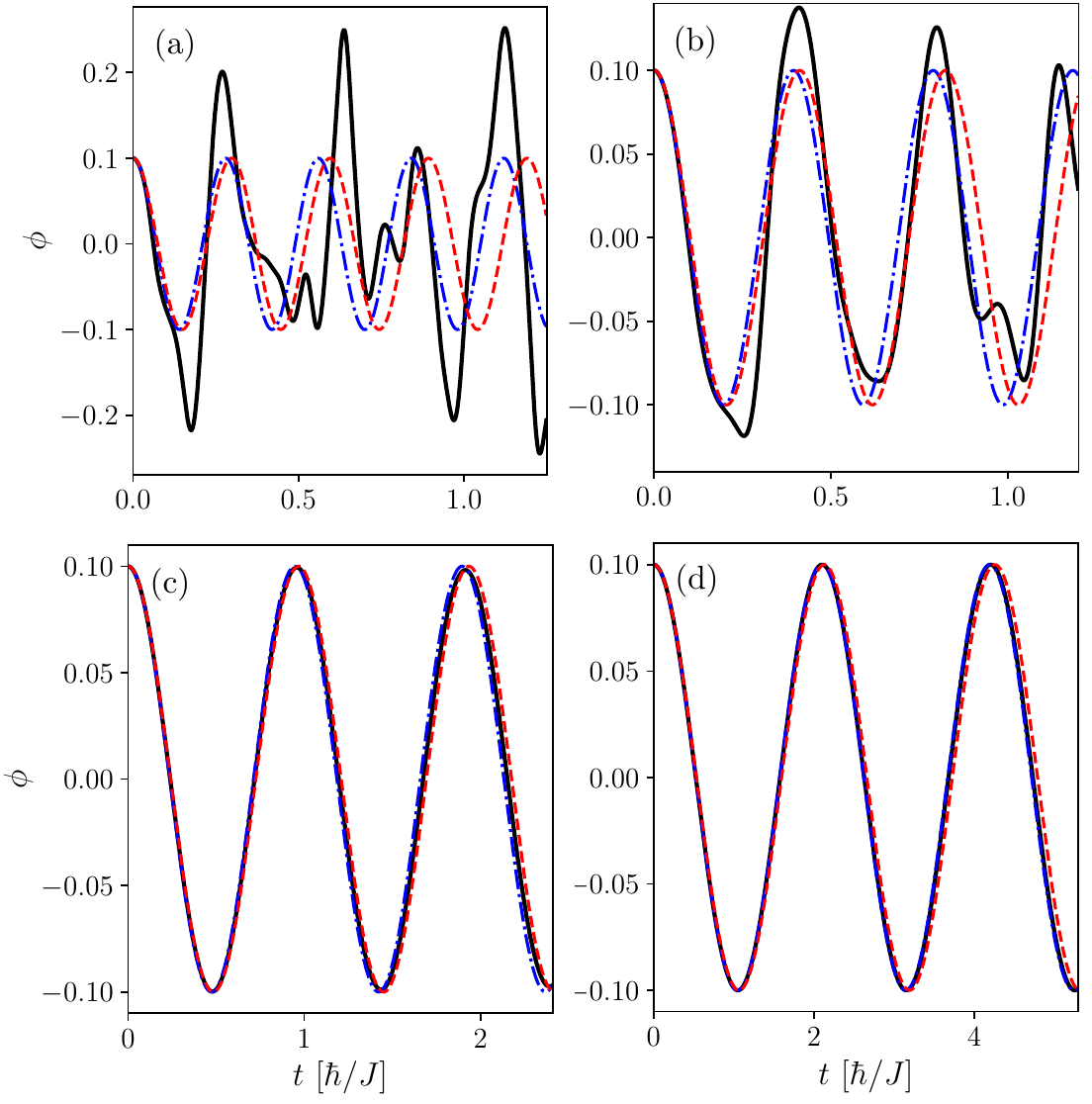}
    \caption{Comparison between the exact dynamics (solid black line), the mean-field dynamics (dashed-dotted blue line), and the quantum-corrected dynamics (dashed red line) of the relative phase, for $N=50$, $J=1.0$, $\phi_0=0.1$, and $z_0=0$, for several values of $U$ in the regions of parameter space where the quantum-corrected dynamics loses accuracy. (a) $U=5.0$, Region III. (b) $U=2.5$, boundary between Region I and Region III. (c) $U=0.4$, boundary between Region I and Region II. (d) $U=0.05$, Region II. (Units: $\hbar=1$).}
    \label{fig4}
\end{figure}

We investigated the range of validity of the quantum-corrected dynamics as a function of $U/J$ and $N$, identifying the region in parameter space where it provides a closer agreement with the exact results than the mean-field approximation. This domain, referred to as Region I in Fig. \ref{fig3}, was determined qualitatively by examining the limits in which the quantum-corrected evolution continues to capture the salient features of the exact evolution, notably the oscillation frequency, over at least three periods. Decreasing $U/J$ and increasing $N$ (while remaining within Region I) reduces many-body modulations and allows the maintain quantitative agreement for longer times. Fig. \ref{fig4} shows representative examples of the dynamics at the boundaries of the validity domain as well as in the parameter regimes where the quantum-corrected results become inaccurate (Regions II and III).

At small $U/J$ (Region II), the quantum-corrected dynamics loses accuracy due to the breakdown of the assumption underlying the Gaussian integration of $z$ (Figs. \ref{fig4}c and \ref{fig4}d), as anticipated in the discussion following Eq. \eqref{wilson}. We find that the transition line between Regions I and II fits $\Lambda = 10$ for $N \gtrsim 25$. However, the mean field provides a quite accurate description of Region II, and in the special case $U=0$, $\phi_\text{cond}(t)$ follows exactly Eqs. \eqref{eqss} at arbitrary $N$. We remark that this result still depends on the choice of $|\Psi_{\phi_0, z_0}\rangle$ as initial state; if we initialized the system in a Fock state, for example, its quantum evolution could certainly not be described by a single mean-field trajectory, because a Fock state has no well-defined relative phase. Instead, there are operators whose quantum evolution (as far as only first moments are considered) is described exactly by the mean-field equations for any initial state. This is due to the fact that for $U=0$ the two-site Bose-Hubbard Hamiltonian can be written as $\hat H = -2J \hat S_x$, where $\hat S_x = \frac{1}{2}(\hat a_L \hat a_R + \hat a_R \hat a_L)$, $\hat S_y = \frac{1}{2i}(\hat a_L^\dag \hat a_R - \hat a_R^\dag \hat a_L)$, and $\hat S_z = \frac{1}{2}(\hat N_L-\hat N_R)$ are spin $N/2$ operators \cite{Leggett}. The unitary operator $e^{-i\hat Ht/\hbar} = e^{i(2Jt/\hbar)\hat S_x}$ is therefore a $SU(2)$ rotation around the $\mathbf e_x$ axis, and for any operator that is a linear combinations of the generators, $\hat O = \mathbf a\cdot \hat{\mathbf S}$, the Heisenberg equation $\partial_t\hat O(t) = -\frac{i}{\hbar}[\hat O, \hat H] = -2J\mathbf a\cdot (\hat{\mathbf S}\times \mathbf e_x)$ are also linear. The expectation value $\langle \hat O(t)\rangle$ on any initial state thus obeys the same linear equation obtained by replacing the operators by classical variables, $\partial_t\langle\hat O(t)\rangle = -2J\mathbf a\cdot (\langle\hat{\mathbf S}\rangle\times \mathbf e_x)$. This is a realization of the Ehrenfest theorem. Finally, it is worth mentioning that deep in Region II, i.e. for $\Lambda \sim 1$ and small $N$, the agreement between the mean-field and the exact dynamics can be improved by introducing the finite-size correction $U \to U(1-N^{-1})$ in Eqs. \eqref{eqss} \cite{Wimberger}.

At large $U/J$ (Region III), the dynamics of $\phi_\text{cond}(t)$ is highly anharmonic and the oscillations' amplitude is strongly modulated on time scales comparable with the inverse of the Josephson frequency. This is a truly quantum regime, where the system's coherence is reduced due to relatively strong interactions (for instance, $\langle\varrho_0(t)\rangle_\mathrm{time}/N \simeq 0.78$ for $N=50$ and $U/J=5$) and neither the mean-field nor the quantum-corrected dynamics provide an accurate description (Figs. \ref{fig4}a and \ref{fig4}b). The separation line between Regions II and III can be set schematically at $U/J=1$ for $N \lesssim 25$, which identifies the transition from a weakly interacting and highly coherent system, which is well described by the mean field theory, to a strongly interacting system displaying genuine many-body effects and reduced coherence.

Josephson oscillations of the relative phase have been directly observed across several platforms, where interference measurements enable precise tracking of coherent phase dynamics between two weakly coupled modes \cite{Albiez, Lagoudakis, Abbarchi, Valtolina}. Given the current capabilities in controlling tunneling rates, interaction strengths, and atom numbers in bosonic Josephson junctions, it appears experimentally feasible, though technically demanding, to test the theoretical prediction of a reduction in the Josephson oscillation frequency arising from quantum corrections. With dedicated effort aimed at achieving high phase coherence and sub-percent frequency precision, such measurements could provide a direct probe of mesoscopic quantum fluctuations in collective condensate dynamics. However, we emphasize that there already exist indirect evidences of the quantum nature of $\phi$. In superconducting Josephson junctions, this was demonstrated decades ago in pioneering experiments by Clarke and collaborators, who observed macroscopic quantum tunneling and energy-level quantization of the phase \cite{Clarke1, Clarke2}. For bosonic Josephson junctions, the condensate fraction (or equivalently, the coherence visibility) constitutes a similarly relevant observable for probing the quantum character of the phase. Within mean-field theory, the ground state of the system always has full condensation, $\varrho_0/N=1$, irrespectively of the interaction strength $U$. Indeed, since in mean-field theory the condensate occupation is given by $\varrho_0/N=(1+\sqrt{1-z^2}\cos\phi)/2$, the minimum of the energy \eqref{ener} at $\phi=z=0$ yields $\varrho_0/N=1$. In contrast, quantum fluctuations introduce a finite zero-point energy that depends on $U$ through the Josephson frequency $\omega_J$. The consequence is a reduction of the condensate fraction in the ground state (quantum depletion) \cite{Vianello}. Incorporating the quantum-corrected frequency $\Omega_J$ improves the agreement with fully quantum calculations of this depletion. Quantum depletion has already been observed in several ultracold-atom experiments (see e.g. Refs. \cite{Xu, Chang, Lopes}) and its precise measurement in a double-well geometry would provide an indirect test of the predicted quantum corrections to the Josephson frequency.

\section{Conclusion}

In this paper, we investigated quantum corrections to the mean-field dynamics of the relative phase between two coupled condensates. Starting from the classical action expressed in the collective variables $(\phi, z)$, we obtained an effective action depending only on the phase, and we derived the corresponding one-loop quantum effective action by means of a covariant background-field method that fully accounts for the spatial dependence of the mass and the exact form of the potential. Comparison with the exact dynamics of the two-site Bose-Hubbard model shows that the one-loop corrected dynamics reproduces the quantum evolution significantly better than the standard mean-field equations over a wide region of the parameter space $(N,\,U/J)$, confirming that the field-theoretic quantum effective action approach provides a systematic and physically transparent framework to capture leading-order quantum effects in Josephson junctions. For small oscillations around the equilibrium value zero, the dynamics is harmonic, as in the mean-field case, but with a modified Josephson frequency. The shift of the classical frequency induced by quantum fluctuations lies between $1\%$ and $3\%$ for the parameter range considered in Fig. \ref{fig3}. 

Besides elucidating the correspondence between the quantum theory and its classical limit, and quantifying the importance of quantum effects for different interaction strengths and number of particles, the corrected Josephson frequency may be employed to enhance the accuracy of semiclassical schemes that can be used in the description of thermodynamic properties of these systems \cite{Vianello}. Future work could generalize the quantum effective-action approach to the full set of collective variables $(\phi, z)$, thereby allowing a systematic study of quantum corrections to phenomena such as macroscopic quantum self-trapping, and could extend the present treatment to finite temperature. 

\bmhead{Author Contributions} L.S. initiated the project. S.S., C.V. and L.S. developed analytical calculations and C.V. performed numerical and graphical tasks. All authors contributed to the writing of the manuscript and the discussion of the results. All authors reviewed the manuscript.

\bmhead{Funding} C.V. and L.S. are supported by “Iniziativa Specifica Quantum” of INFN and by the Project “Frontiere Quantistiche” (Dipartimenti di Eccellenza) of the Italian Ministry of University and Research (MUR). L.S. is partially supported by funds of the European Union-Next Generation EU: European Quantum Flagship Project “PASQuanS2”, National Center for HPC, Big Data and Quantum Computing [Spoke 10: Quantum Computing]. L.S. also acknowledges the PRIN Project “Quantum Atomic Mixtures: Droplets, Topological Structures, and Vortices” of MUR.

\bmhead{Data Availability} Data are available from the authors upon reasonable request.

\bmhead{Competing Interests} The authors declare no competing interests.

\bibliography{References}

\end{document}